\documentclass[letterpaper]{article}
\usepackage{prepr}

\usepackage{amsmath,amsfonts} 
\usepackage{url}
\usepackage{booktabs} 
\usepackage{graphicx}
\usepackage{textcomp}
\usepackage{xcolor}
\usepackage{xspace}
\usepackage{hyperref}
\usepackage{balance}
\usepackage{colortbl}
\usepackage{tabularx}
\usepackage{tabulary}
\usepackage{booktabs}
\usepackage{multicol}
\usepackage{multirow}
\usepackage{rotating}
\usepackage{caption}
\usepackage{subcaption}
\usepackage[export]{adjustbox}
\usepackage{tabularx}
\usepackage{tikz}
\usepackage{float}
\usepackage[shortlabels]{enumitem}
\usepackage[most]{tcolorbox}
\usepackage{xparse}
\usepackage{etoolbox}
\usepackage{xurl}
\usepackage{tabularx}
\usepackage{siunitx}

\usepackage{varioref}
\usepackage{array}
\usepackage{ragged2e}
\usepackage[capitalise]{cleveref}
\usepackage{scalerel}

\crefname{sublisting}{Listing}{Listings}
\Crefname{sublisting}{Listing}{Listings}
\Crefname{equation}{Eq.}{Eqs.}
\Crefname{figure}{Fig.}{Figs.}
\Crefname{tabular}{Tab.}{Tabs.}
\Crefname{section}{Sec.}{Secs.}

\definecolor{commentGreen}{RGB}{128,127,41}
\definecolor{stringGreen}{RGB}{0,127,38}
\definecolor{keywordRed}{RGB}{151,0,11}
\definecolor{typePurple}{RGB}{128,3,82}
\definecolor{variableBlue}{RGB}{0,0,255}
\definecolor{annotationRed}{RGB}{198,4,38}
\definecolor{annotationBlue}{RGB}{32,21,223}

\DeclareMathAlphabet{\mathcal}{OMS}{cmsy}{m}{n}
\newcommand\Csh{C\scalerel*{\#}{X}}

\definecolor{01gen}{RGB}{245,245,245}
\definecolor{02los}{RGB}{213,232,212}
\definecolor{03act}{RGB}{225,213,231}
\definecolor{04ref}{RGB}{218,232,252}
\definecolor{05misc}{RGB}{255,242,204}

\def\@keywords#1,#2\@nil{%
  #1%
  \if\relax\detokenize{#2}\relax
  \else
    ~\textbullet~%
    \@keywords#2\@nil
  \fi
}

\newtcolorbox[auto counter]{reqbox}[1]{title={\bfseries #1},
  coltitle=white, 
  leftrule=0.25mm,
  rightrule=0.25mm,
  bottomrule=0.25mm,
  toprule=0.25mm,
  colframe=white!45!black, boxsep=2pt,left=4pt,right=4pt,top=3pt,bottom=3pt,  sharpish corners}
\usepackage{tabularx}
\usepackage{booktabs}
\title{A Model-Driven Automation Pipeline for Building Operation}

\author{
\PREPauthor{Elias Kohler}{University of Innsbruck, Department of Computer Science}{elias.kohler@student.uibk.ac.at}
\PREPauthor{Sascha Hammes}{University of Innsbruck, Unit of Energy Efficient Building}{sascha.hammes@uibk.ac.at}
\PREPauthor{Michael Vierhauser}{University of Innsbruck, Department of Computer Science}{michael.vierhauser@uibk.ac.at}
\PREPauthor{Philipp Zech}{University of Innsbruck, Department of Computer Science}{philipp.zech@uibk.ac.at}
}

\setvenue{\textbf{27th International Conference on Product-Focused Software Process Improvement}}
\setvenueone{the \textbf{27th International Conference\\on Product-Focused Software Process Improvement (PROFES)}}
\setyear{2026}
\setdoi{}

\begin{document}
\pagestyle{custom}
\maketitle

\begin{abstract}
Building Control Systems (BCS) engineering in the Architecture, Engineering, Construction~\& Operations (AECO) domain remains a complex, manual, and error-prone process, with fractured tools and manual commissioning causing design and built reality to diverge into a performance gap. We argue that a fully integrated and automated, tool-supported workflow spanning design, programming, and operation is feasible and can improve efficiency while reducing manual effort and configuration errors. Building on model-driven engineering, we present an automated process pipeline for modeling, programming, and auto-commissioning BCS. It uses two domain-specific modeling languages with tooling support to synthesize and deploy executable artifacts from platform-independent models, replacing a fractured, multi-stakeholder handoff process with a defined four-step workflow and improving consistency across the engineering lifecycle. Our evaluation reports notable efficiency gains, improved adaptability, moderate-to-high user acceptance, and good performance with solid scalability.
\end{abstract}

\section{Introduction}
\label{sec:introduction}

Over the last decades, Software Engineering (SE) has played a central role in enabling automation in a wide variety of application domains~\cite{ajiga2024role}. In turn, SE has also tremendously benefited from advances in automation~\cite{liu2019automatic}. Model-Driven Engineering (MDE) is a prime example of this relationship, capitalizing on software tools and formal models to provide higher-level abstraction and manage complexity, while simultaneously enabling automation of various process activities along the software development lifecycle (SDLC)~\cite{whittle2013state,hutchinson2011model,kent2002model}. 
To date, despite the application for many engineering activities, MDE however remains largely confined to SE, and parts of systems engineering and the domain of Cyber-Physical Systems (CPS)~\cite{mohamed2021model}. 

However, even in the area of CPS engineering, MDE has not yet touched on all aspects, and significant gaps remain in the end-to-end automation of development workflows and processes. For example, in building engineering~\cite{rajkumar2010cyber}, the engineering of Building Control Systems (BCS) remains a fractured, largely manual, barely automated, and often error-prone task~\cite{hauer2024integrating,hammes2024market}. While Building Information Modeling (BIM) already provides a model-based approach for \emph{virtually integrated design, construction, and operation} (ViDCO) that comprises object-based modeling, model-based collaboration, and network-based integration~\cite{succar2012measuring}, BCS engineering still remains barely automated. This is further exacerbated by the lack of proper tool support and by existing media breaks at the modeling level~\cite{hammes2024market}. These issues become evident during information exchange between both stakeholders and their tools and systems, respectively~\cite{hammes2024market,hauer2024integrating,zech2025agile}.

Both MDE and BIM emphasize model-centric abstractions and development, separation of concerns, and potential end-to-end automation from design models to deployed systems, meriting deeper investigation for enabling an integrated model-driven workflow and streamlined process for BCS engineering, from design to operation. Such a workflow facilitates control logic development by elevating automation programming from low-level, vendor-specific code to high-level, domain-specific models, thereby systematically reducing the complexity of managing heterogeneous devices and languages while increasing automation throughout the development lifecycle. We hypothesize that such a workflow can yield benefits consistent with MDE in software and system engineering, including improved efficiency, ease of use, and consistency~\cite{DANIEL2024103032,JAMES2024103057,BUSCHHAUS2024103033,NISTALA2024103058}. 

To address these challenges, in this paper, we introduce a novel model-driven tool environment that realizes an end-to-end automation pipeline for integrated BCS modeling, programming, and commissioning. Our approach is based on two Domain-specific Modeling Languages (DSMLs)~\cite{wkasowski2023domain} \emph{that enable the specification of device configurations and control logic}. Using model-to-text (M2T) generation~\cite{kahani2019survey}, the pipeline emits runtime artifacts (device configurations and control logic) which are automatically deployed to a physical BCS installation, thereby exploiting model-based tool integration (MBTI)~\cite{kapsammer2006model} to connect development tools and system hardware.
Our evaluation reports notable efficiency gains and improved system adaptability, moderate-to-high user acceptance, and good performance with solid scalability due to the framework's end-to-end automation and high-level, platform-independent modeling capabilities. As part of this, we claim the following major contributions:\vspace{-0.1em}
\begin{description}
    \item \textbf{-- Two DSMLs} for platform-independent specification of BCS hardware installations and control logic for clear separation of concerns between physical device topology and behavioral logic.
    \item \textbf{-- An integrated model-driven development and deployment pipeline} that combines DSML-based modeling, automated validation, code generation, and deployment to provide an end-to-end automated workflow from high-level (vendor-independent) models to operational (vendor-specific) BCS installations.
    \item \textbf{-- A systematic modeling methodology} comprising four steps that supports both initial system commissioning and rapid post-commissioning adaptations.
    \item \textbf{-- A defined four-step engineering process} with explicit role assignments that replaces the incumbent fractured handoff chain, supporting both initial commissioning and incremental post-commissioning adaptations without re-engaging an executing company.
    \item \textbf{-- An evaluation} demonstrating notable efficiency gains regarding commissioning, improved system adaptability, moderate-to-high user acceptance, and good performance with solid scalability in a mobile demo setting with sensors and LED lights using Beckhoff automation hardware (CX5140 controller).
    \item \textbf{-- A fully implemented and publicly available artifact} including language definitions, code generators, and deployment tools for the Beckhoff TwinCAT 3 platform~\cite{artifact}.
\end{description}
The remainder of this paper is structured as follows. \cref{sec:problem} outlines our research problem, followed by \cref{sec:solution}, our automated model-driven process pipeline. \cref{sec:evaluation} evaluates our work, and \cref{sec:discussion} then discusses our results before we conclude in \cref{sec:conc}.

\section{Background and Motivating Example}
\label{sec:problem}

In the following, we outline the research problem centered around the \emph{fractured communication and information flows} among stakeholders involved in BCS engineering that arise from \emph{isolated and largely manual processes}. This fragmentation leads to a significant gap between the designed and the actual operational performance of buildings~\cite{zech2025bs}, a crucial aspect addressed in this paper by capitalizing on MDE.

\subsection{Motivating Scenario}
\label{ssec:scenario}

\begin{figure}[bth]
    \centering
    \includegraphics[width=.95\linewidth]{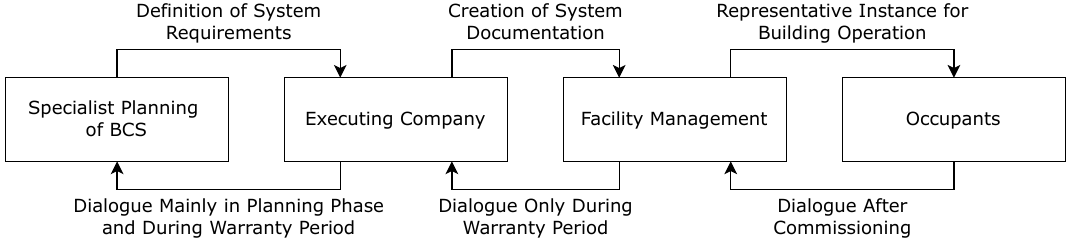}
    \caption{The fragmented BCS engineering communication chain, highlighting siloed interactions and disconnected feedback loops.}
    \label{fig:problem}
\end{figure}

Imagine a common scenario in building automation, where office floor lighting uses a closed-loop system for automated light level adjustment. 
Following the fractured, yet common process depicted in \cref{fig:problem}, a \emph{specialist planner} specifies this logic, which an \emph{executing company} then manually programs into a controller. This unstructured handover and manual translation is where errors and a lack of transparency creep in~\cite{hammes2024market}.

However, the real problems often surface much later, during operation. During normal operation, occupancy patterns and lighting requirements change over time. For example, shifts in occupancy mean spaces are used differently than originally intended, rendering preset system configurations outdated. Workplace reorientations may also change the conditions for daylight control. Existing shading systems may not accommodate such changes, potentially resulting in glare. To restore comfort and energy efficiency, the facility manager must identify affected zones, determine the required parameter changes, and reconfigure both the artificial lighting and the shading systems. These interventions are time-consuming, error-prone, and costly, a problem further compounded by the limited degree of automation typically found in such systems. This epitomizes the system's limited ability to adapt, a direct result of the siloed and fractured approach~\cite{hammes2024market,shen2014energy}.

\begin{figure}[bth]
    \centering
    \includegraphics[width=.9\linewidth]{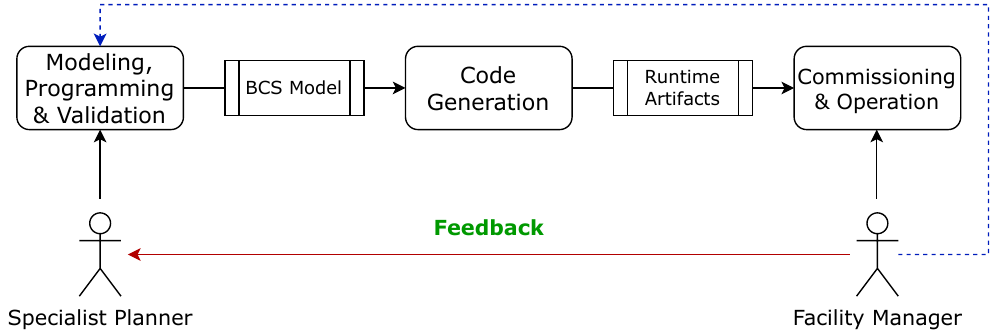}
    \caption{The proposed unified, model-driven workflow, showing stakeholders in an integrated tool-chain for end-to-end automation from modeling to operation (cf.~\cref{fig:problem}).}
    \label{fig:solution}
\end{figure}

Our proposed process (cf.~\cref{fig:solution}) addresses these challenges by providing an automated, model-driven workflow where the \emph{specialist planner} captures the logic in a formal BCS model, which acts as the single source of truth for the subsequent automated code generation and commissioning process. When the space utilization requirement changes, the \emph{facility manager} no longer needs to rely on an \emph{executing company}; instead, he/she can directly trigger changes to the BCS model with subsequent automatic regeneration and redeployment. This replaces manual, error-prone re-commissioning with an automated workflow that is quick, traceable (the text-based nature of our DSMLs renders models versionable, e.g., via Git), and reliable, thereby allowing building operations to adapt iteratively and in an agile way as building occupancy and usage evolves.

\subsection{Related Work}

Our work sits at the intersection of MDE and the AECO domain, the emerging area of \emph{Model-driven Engineering for Building Operations} (MDE4BO). 
The review by Götz et al.~\cite{gotz2020model} frames our context. It notes that MDE-for-AECO is discussed mainly in domain venues rather than SE conferences, and calls for work on unified modeling and models@run.time~\cite{nelly19}. We aim to address this by enabling model-driven adaptation during a building's operational phase. 

Stratbücker et al.~\cite{stratbucker2017model} use ATL-based model-to-model (M2M) transformations to feed simulation tools (e.g., EnergyPlus or Modelica) to target interoperability rather than executable control, as also investigated by Zech et al.~\cite{zech2025agile}. Schachinger and Kastner~\cite{schachinger2015model} generate Web Service gateway configurations from platform-independent models, but focus on integration, not control logic. Knorr et al.~\cite{knorr2024model} transform BIM models into UML and generate both simulation models and deployable Structured Text (ST), but rely on general-purpose UML, which adds manual overhead. Cai et al.~\cite{cai2023automated} generate PLC code but require programmatic model creation and manual tuning, limiting rapid adaptation. These works share our use of MDE for interoperability, but target design and analysis artifacts. Our work synthesizes deployable control logic and device configurations for commissioning and operation.

Kirchhof et al.~\cite{kirchhof2022montithings} present MontiThings for reliable IoT applications with automated deployment. Corradini et al.~propose FloWare~\cite{corradini2023floware} which pairs feature models with low-code platforms. These IoT platforms, however, are general-purpose and target different runtimes than BCS, whereas we target the specific workflow of BCS engineering and industrial PLCs as runtime environment.

\subsection{Research Problem}

Current MDE approaches for BCS usually focus on generating models for analysis and simulation, while frameworks in the broader IoT domain are not tailored to the specific stakeholders, workflows, and target platforms of BCS engineering. This leaves a clear gap: no unified, vendor-independent workflow exists that bridges high-level design to operational control by integrating platform-independent modeling using DSMLs with a model-driven generation and deployment pipeline for industrial PLCs in BCS. Existing process standards for the AECO domain, such as ISO~19650~\cite{ISO19650-1}, govern information management across the building lifecycle but do not address the engineering of BCS, leaving the process gap we target unaddressed. To address this gap, we frame our research problem using Design Science Research (DSR)~\cite{wieringa2014dsr}: the construction and evaluation of a novel technical artifact for model-driven BCS engineering. We formally state our design problem as follows and derive three research questions (RQ):
\begin{table}[H]
    \scriptsize
      \begin{tabularx}{\columnwidth}{lX}
        \toprule
        \textbf{Improve} & \emph{building Control Systems engineering} (context) \\
        \textbf{by designing} & \emph{a model-driven process pipeline that decouples system design from implementation} (artifact) \\
        \textbf{that satisfies} & \emph{platform independence, hardware/logic separation, and automation} (requirement) \\
        \textbf{in order to} & \emph{reduce engineering complexity, improve adaptability, and help close the energy performance gap}. (goal) \\
        \bottomrule
    \end{tabularx}
\end{table}


\textbf{RQ1:} How can MDE, utilizing DSMLs, enable the creation of abstract, platform-independent hardware and control models that can be transformed into vendor-specific implementations to facilitate intuitive system definition and automated deployment?

\textbf{RQ2:} What is the impact of a model-driven process on the efficiency, adaptability, and automation of BCS engineering, considering factors such as development time, maintenance effort, and system responsiveness to changing requirements? 

\textbf{RQ3:} What are the resulting benefits and implications in terms of acceptance, perceived usefulness, and ease of use when using our proposal for BCS engineering?\\ 

The remainder of our paper answers these RQs through the construction and evaluation of a prototype artifact for model-driven engineering of BCS. We discuss the architecture and implementation of our framework and present a comprehensive evaluation that is grounded in a study that targets a Beckhoff\footnote{https://www.beckhoff.com} automation system to demonstrate the practical feasibility of our proposal.

\section{Solution Overview}
\label{sec:solution}

Based on the previously described research problem, we define five core requirements derived from (i) identified gaps in current BCS engineering practice~\cite{hammes2024market,hauer2024integrating}, (ii) established MDE principles~\cite{hutchinson2011model,kent2002model}, and (iii) the specific needs of AECO domain stakeholders~\cite{gotz2020model,hammes2024market,hauer2024integrating}. Requirements are specified in \cref{tab:requirements} using the EARS notation~\cite{mavin2009easy}. 
\begin{table}[bth]
    \scriptsize
    \centering
    \caption{Core requirements derived from BCS engineering gaps, MDE principles, and AECO stakeholder needs, specified using EARS notation~\cite{mavin2009easy}.}
    \label{tab:requirements}
    \begin{tabularx}{\linewidth}{@{}>{\RaggedRight\arraybackslash}p{2.4cm}>{\RaggedRight\arraybackslash}X@{}}
        \toprule
        \textbf{Requirement} & \textbf{Description and Rationale} \\
        \midrule
        \textbf{R1: Abstraction from Platform Complexity}
        & The system shall abstract vendor-specific details and support platform-independent specification of hardware configurations and control logic. \emph{Addresses the fractured tool landscape in BCS engineering~\cite{hammes2024market,hauer2024integrating} and follows established MDE principles of platform-independent modeling~\cite{kent2002model,schachinger2015model,zech:models:2025}.} \\
        \addlinespace
        \textbf{R2: Unified and Holistic Modeling}
        & The system shall enable integrated modeling of hardware topology and control logic within a single, unified modeling framework. \emph{Current approaches usually model these aspects in isolation~\cite{knorr2024model,schachinger2015model}, leading to consistency issues and manual integration effort.} \\
        \addlinespace
        \textbf{R3: Automated End-to-End Workflow}
        & The system shall support an automated workflow that transforms high-level models into executable code and deploys it to target platforms. \emph{Manual translation steps are a major source of errors and inefficiency in current BCS engineering~\cite{hammes2024market}, which automated model-driven generation~\cite{kahani2019survey} eliminates.} \\
        \addlinespace
        \textbf{R4: Accessibility}
        & When users (e.g., \emph{facility managers}) modify control parameters, the system shall support intuitive and safe interaction without requiring deep programming expertise. \emph{BCS adaptations during building operation often require costly re-engagement with executing companies~\cite{hammes2024market,shen2014energy}.} \\
        \addlinespace
        \textbf{R5: Scalability}
        & As model size increases, the system shall generate artifacts efficiently and maintain manageable modeling complexity for users. \emph{Real-world BCS installations comprise hundreds of I/O points and complex control logic~\cite{hauer2024integrating}, which the solution must handle without degrading performance.} \\
        \bottomrule
    \end{tabularx}
\end{table}

\begin{figure}[bth]
    \centering
    \includegraphics[width=.99\linewidth]{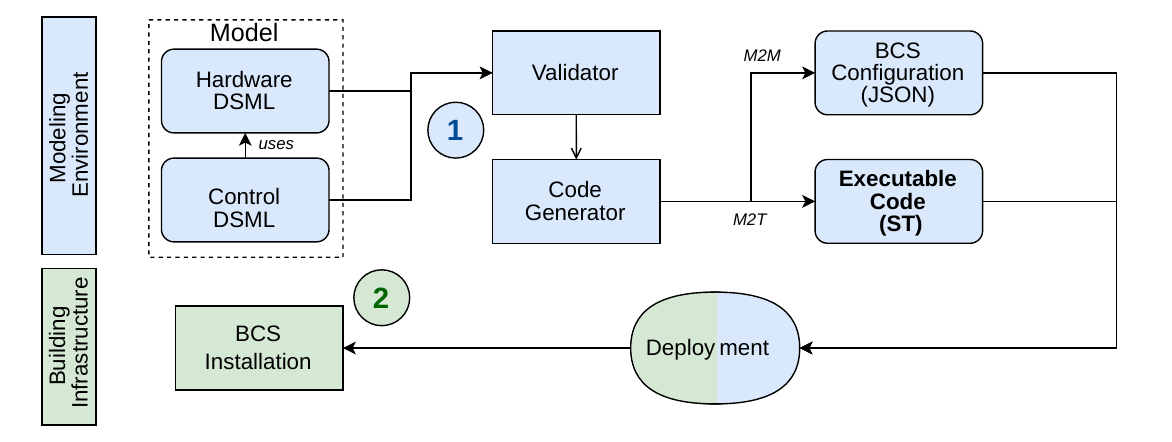}
    \caption{Proposed conceptual architecture following the workflow from \cref{fig:solution}.}
    \label{fig:overview}
\end{figure}

\cref{fig:overview} outlines our solution based on two DSMLs, viz.~the \texttt{Hardware DSML} and the \texttt{Control DSML}. The former is designed to model a BCS installation, while the latter is used for modeling control logic regarding available hardware, such as sensor data (cf.~illuminance), and commands (e.g., dimming of luminaries). Both models are subsequently validated and leveraged for model transformation, viz.~(i) M2M transformation, to emit the BCS configuration in JSON, and (ii) M2T transformation to emit control logic in ST. As a last step, these artifacts (JSON, ST) are deployed to the BCS using vendor-specific deployment tools. As a result, our solution establishes a unified workflow that eliminates manual steps, supports rapid system changes, and uses models as first-class citizens during the engineering process. In the following, we describe our modeling infrastructure and its associated modeling methodology (cf.~\cref{ssec:modeling}), code generation and deployment (cf.~\cref{ssec:gen_dep}), and provide relevant implementation details on information on artifact availability.

\subsection{Modeling Infrastructure and Methodology}
\label{ssec:modeling}

In the following, we provide a comprehensive overview of our modeling infrastructure and methodology, enabling systematic design and update of a BCS and its control logic. The methodology covers the specification of architectural and behavioral models of a BCS. The architectural design of a building however\textemdash which usually precedes integrating a BCS\textemdash is out of scope. The \texttt{Hardware DSML} is used for structural modeling of a BCS, whereas the \texttt{Control DSML} is used for behavior modeling. If needed, syntax and semantics can be adapted/extended. Metamodels and example of both the \texttt{Hardware} and \texttt{Control DSML} are available from our artifact~\cite{artifact}.\vspace{-0.75em}

\subsubsection{Hardware DSML}
\label{ssec:h-dsml}

The \texttt{Hardware DSML} is specifically designed for describing BCS hardware in a platform-independent way. Crucially, its design separates the definition of physical devices from their logical signals. A \emph{physical view} models the real-world hardware, such as fieldbuses, controllers, I/O modules, and their connections. The physical view reflects the actually installed devices and their arrangement, e.g., which bus a module is connected to, the module type, its slot position, and network addresses. In contrast, a \emph{logical view} abstracts collections of I/O signals into named \emph{datapoints} and \emph{port groups}. Logical channels are used by the \texttt{Control DSML} to enable the user to decouple the functional meaning of signals (such as "room temperature sensor" or "presence button") from their physical assignments.

On the \emph{physical} side, the \texttt{Hardware DSML} models the device hierarchy: a \texttt{Bus} (e.g., KBus or EtherCAT) contains one or more \texttt{Box}es (cabinets or terminal blocks), each holding \texttt{Module}s at specific slots, together with \texttt{NetworkSettings} (e.g., controller IP) needed for deployment. On the \emph{logical} side, a \texttt{Controller} groups \texttt{PortGroup}s (sets of I/O channels of a given type and count) and \texttt{Datapoint}s, where each \texttt{Datapoint} maps one or more named logical \texttt{Channel}s, with their datatype (e.g., \texttt{BOOL}, \texttt{REAL}), to concrete physical channels. This logical layer is what the \texttt{Control DSML} models against, thereby decoupling control logic from physical wiring.\vspace{-0.75em}

\subsubsection{Control DSML}
\label{ssec:c-dsml}

The \texttt{Control DSML} is designed to describe high-level control logic for BCS. It builds on the abstraction layer provided by the \texttt{Hardware DSML}, allowing all logic to be written against logical \emph{datapoints}, fully decoupled from physical wiring or vendor-specific interfaces. This separation enables the reuse of control logic across multiple projects and vendors, and simplifies later system modifications. Specifically, the \texttt{Control DSML} enables succinct and unambiguous specification of building control behavior, supports both procedural and declarative logic patterns common in the BCS domain, encourages modularity and reuse through function blocks, and facilitate analysis and automated code generation via a static, grammar-based structure.

A \texttt{ControlModel} is the root container. Its \texttt{ControlBlock} is linked to a \texttt{Controller} from the \texttt{Hardware DSML} and holds global declarations and logic units. Beyond conventional declarations (variables, structs, enums, type aliases, and \texttt{import}/\texttt{extern} for integrating vendor libraries such as Tc3\_DALI), two constructs carry the domain-specific weight. First, a \texttt{FunctionBlock} encapsulates reusable logic with explicit inputs, outputs, and locals, and is instantiated via a \texttt{use} statement that maps program variables to its ports. Second, a \texttt{ControlUnit} organizes executable logic with continuous (\texttt{default}), scheduled (\texttt{at}), or conditional (\texttt{when}) execution, and supports declarative reactions: \texttt{on\_rising}/\texttt{on\_falling} for boolean edge changes and \texttt{after} for on-delay timer logic. These declarative constructs are what the generator rewrites into imperative ST.

\subsubsection{Modeling Steps}
\begin{figure*}[bth]
    \centering
    \includegraphics[width=\textwidth]{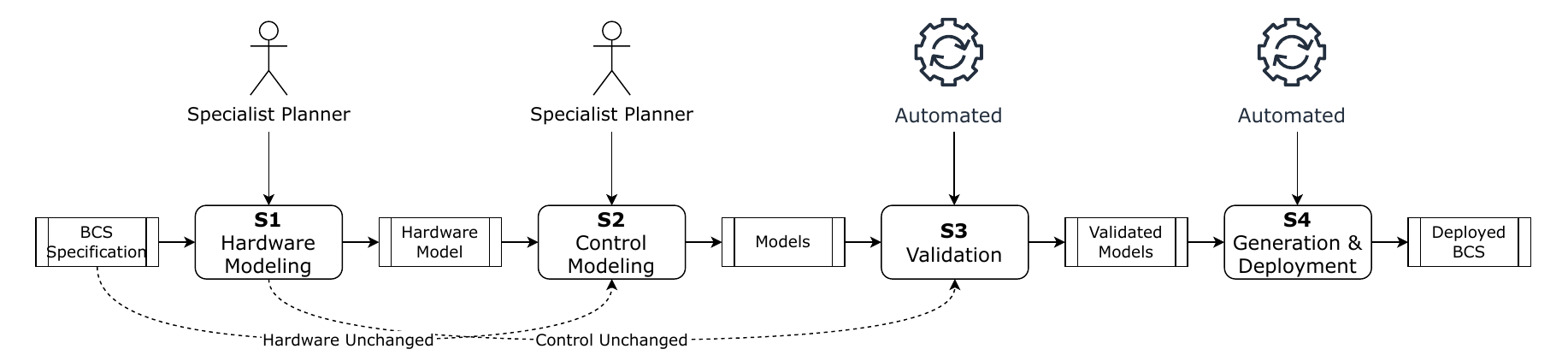}
    \caption{Four-step BCS engineering process. S1 and S2 are manual activities performed by the specialist planner; S3 and S4 are fully automated. Dashed arrows show optional skip paths for post-commissioning adaptations, only affected stages re-run.}
    \label{fig:process}
\end{figure*}

\cref{fig:process} shows the four-step BCS engineering process. In \textbf{S1}, the \emph{specialist planner} captures the physical and logical structure of the BCS installation using the \texttt{Hardware DSML} to produce a hardware model of the BCS as the handoff artifact to S2. In \textbf{S2}, the \emph{specialist planner} models control behavior against the logical datapoints defined in S1 using the \texttt{Control DSML}. \textbf{S3} and \textbf{S4} are fully automated: the pipeline validates both models for completeness, type safety, and mapping consistency, then transforms and deploys the generated artifacts to the BCS without manual intervention. For post-commissioning adaptations, S1 can be skipped if the hardware is unchanged, and S2 can usually be skipped if only hardware parameters change, that is, only affected stages re-run. This directly addresses the re-commissioning bottleneck from \cref{sec:problem} where the facility manager triggers a re-run from whichever stage is affected, without re-engaging an executing company.

\subsection{Generation and Deployment}
\label{ssec:gen_dep}

After successful validation, the hardware and control models serve as input to code generation and deployment (cf.~\cref{fig:overview,fig:process}). Our code generator follows a template-based, visitor-pattern approach by traversing the abstract syntax trees of both models and populating platform-specific templates with information from both models~\cite{felderer2009towards}. Each pattern is responsible for translating a specific model construct (cf.~\cref{ssec:h-dsml,ssec:c-dsml}) into a platform-specific artifact. In our implementation, we have used the Beckhoff TwinCAT 3 system~\cite{beckhoff_automation_interface} as the target BCS platform, which expects configurations in JSON and control logic in ST.

Once all artifacts are generated, an automated deployment tool initializes or updates the BCS. For Beckhoff TwinCAT, the deployment process is orchestrated via TwinCAT's Automation Interface, which handles creation, code upload, I/O linking, and activation. This fully automated pipeline ensures that changes in the high-level models are rapidly and reliably reflected in the deployed BCS solution, with minimal manual intervention and maximal traceability. 

\subsection{Implementation Details}

Both DSMLs are defined using Eclipse Langium~\cite{eclipse_langium} within VS Code. In addition, code generation leverages Langium for model processing and parsing. In our current work, the deployment tool is implemented using \Csh/.NET. Our solution supports a fully automated workflow, from high-level model editing to executable system deployment. Our solution artifact is available online~\cite{artifact}.

\section{Evaluation}
\label{sec:evaluation}

We evaluate our approach through empirical studies that address \textbf{RQ2} and \textbf{RQ3}. \textbf{RQ1} is addressed through the framework design and implementation presented in \cref{sec:solution}, which demonstrates how DSMLs enable platform-independent modeling and automated transformation to vendor-specific implementations.
For \textbf{RQ2}, we apply the Goal-Question-Metric (GQM) paradigm~\cite{caldiera1994goal} to systematically assess the quantitative and qualitative impacts on process efficiency, adaptability, and automation. For \textbf{RQ3}, we conduct a user study employing the Technology Acceptance Model (TAM)~\cite{davis_user_1989,riemenschneider2001explaining} to evaluate acceptance, perceived usefulness, and ease of use from the perspective of domain practitioners. Both evaluations are grounded in a mobile testbed using Beckhoff automation hardware, which serves as our experimental platform. This physical testbed comprises a CX5140 controller, one KL1404 digital input terminal, and one KL2404 digital output terminal to control a lighting system (emulated via LEDs). This setting provides a realistic BCS engineering context and allows us to observe both initial commissioning and post-occupancy adaptation scenarios.

\subsection{Efficiency, Adaptability, and Automation}
\label{ssec:efficiency}

Our goal is to assess the proposed model-driven process with respect to its efficiency, adaptability, and automation in BCS engineering. We thus decompose \textbf{RQ2} into three specific evaluation questions:

\begin{description}[leftmargin=0em]
    \item[\textbf{EQ1: Engineering Effort.}] To what extent does the model-driven process reduce engineering effort and time compared to a traditional manual process during initial system design and commissioning?
    \item[\textbf{EQ2: System Adaptability.}] To what extent does the model-driven process improve efficiency when adapting an existing BCS to changed building requirements?
    \item[\textbf{EQ3: Low-Barrier Automation.}] Does the DSML enable engineers without PLC background to implement building automation functions with minimal guidance?
\end{description}
\cref{tab:gqm-metrics} maps these evaluation questions to concrete metrics collected during our study.
\begin{table}[bth]
    \scriptsize
    \centering
    \caption{GQM mapping: evaluation questions and metrics}
    \label{tab:gqm-metrics}
    \begin{tabular}{p{.5cm}p{2.3cm}p{4.9cm}}
        \toprule
        \textbf{EQ} & \textbf{Metric} & \textbf{Description} \\
        \midrule
        EQ1 & Time-on-task (initial) & Time from requirements to deployed system \\
        EQ1 & Engineering effort & Lines of code/configuration written \\
        EQ1 & Tool interactions & Number of tools and manual steps required \\
        \midrule
        EQ2 & Time-on-task (update) & Time to modify and redeploy system \\
        EQ2 & Change propagation & Number of artifacts that must be manually updated \\
        \midrule
        EQ3 & Functional correctness & Whether participants produced working automation logic \\
        EQ3 & Hint dependency & Number of sessions requiring external guidance \\
        \bottomrule
    \end{tabular}
\end{table}
We used a within-subjects design in which participants completed three tasks using our model-driven process with the \texttt{Hardware} and \texttt{Control DSML}s. The three tasks were counterbalanced to mitigate learning effects. Participants ($n=15$) comprised 8 experts in building services engineering, 1 energy planner, and 6 computer scientists. Regarding prior DSML experience, 5 computer scientists and 1 civil engineer reported familiarity with DSMLs. Regarding PLC programming experience, 1 computer scientist reported prior experience, 4 experts in building services engineering reported basic experience. The remaining participants reported no experience in either (DSML and PLC programming, respectively). All participants conducted the following tasks:
\begin{description}[leftmargin=0em]
    \item[\emph{S1: Morning Lights.}] A task requiring participants to define a simple push button and time-based lighting control. This scenario was designed to introduce the participants to the DSML, including the use of datapoints and scheduled control.
    \item[\emph{S2: Occupancy-based Heating.}] In a more complex task, participants had to implement a function block that calculates the setpoint temperature based on the current heating mode, which is then used to set an analog output. The heating mode should fall back to ’ECO’ once the room remains unoccupied for a given amount of time.
    \item[\emph{S3: Hardware Adaptation.}] This scenario focused on adaptability. Participants were given the working solution from S2 and were required to perform a small retrofitting task to support an additional room. This required first replacing a digital input terminal in the hardware model, then defining new data point links for the button and luminaire, and finally implementing the luminaire toggle logic.
\end{description}
For each task, we recorded the metrics from \cref{tab:gqm-metrics}. Tasks were considered complete when the system operated correctly in the physical test bed.  As a baseline, we compared the measured efforts with a manual implementation in TwinCAT (and vendor tools) performed by one of the authors.

For \textbf{EQ1}, as a baseline, one author implemented all three scenarios manually in TwinCAT, including documentation review, taking approximately 3 hours. The DSML removes this burden through constrained syntax and built-in validation that guide users toward correct implementations without platform knowledge. All 15 participants, including those with no prior DSML or PLC experience, completed the three scenarios within roughly 45 minutes, and the DSML consistently produced 31--37\% fewer lines of code than the baseline across S1--S3. This substantially reduces both implementation time and code volume, with gains likely larger still for teams lacking deep PLC expertise.

\textbf{EQ2} targets the hardware adaptation scenario (S3) where participants extended the working S2 solution to support a new room. With the DSML, all changes localized to two places, e.g., one module declaration in the \texttt{Hardware DSML} and the corresponding \texttt{Control DSML} logic, with an average adaptation time under 15 minutes, against approximately 30 minutes for the manual baseline. Although the number of artifacts touched was similar, the manual approach required navigating TwinCAT's specific I/O linking and knowing where changes propagate. As the models are decoupled, a physical change in one does not require hunting for dependencies in the other. The resulting $\sim$2$\times$ speedup suggests the DSML meaningfully improves adaptability, especially for post-occupancy changes where speed and low error risk matter most.

As to \textbf{EQ3}, out of 15 sessions, 14 produced correct implementations and the one unsuccessful attempt still yielded a partial solution. Hints were used in most sessions, but were typically small and targeted, e.g., pointing to a specific construct like \texttt{on\_rising}, \texttt{after}, or the concept of a \texttt{unit}. No participant needed step-by-step guidance. Given the participants' varying backgrounds, this indicates the DSML does not require deep technical expertise (cf.~\cref{ssec:acceptance}). The main friction points were syntactic (missing semicolons or keywords, unfamiliar \texttt{enum} notation) rather than conceptual. Several participants noted that auto-formatting and editor support helped. Overall, the results support our claim that the DSML enables engineers without a PLC background to implement building automation functions with minimal guidance.

\subsection{User Acceptance and Perceived Usefulness}
\label{ssec:acceptance}

To address \textbf{RQ3}, we conducted a user study grounded in the TAM~\cite{davis_user_1989,riemenschneider2001explaining}, which posits that perceived ease of use (EOU) and perceived usefulness (USF) drive behavioral intention to use a technology (USE), with training and prior experience (TRA) as external antecedents. We operationalized the model using 14 items across four constructs (TRA, EOU, USF, USE) on a 7-point Likert scale, and the same 15 participants from \cref{ssec:efficiency} completed the survey after all three tasks. We estimated a PLS-SEM model with bootstrapping using R~(v4.4.2) and the \texttt{seminr} package~\cite{hair2021seminr,hair2021partial}. Reliability was satisfactory across all constructs ($\rho_C > 0.70$), and discriminant validity held (all HTMT $< 1.0$). A few items loaded below the 0.708 threshold (most notably SEM.USF.7), but we retained them as each captures a conceptually relevant aspect of its construct (SEM.USF.7, for instance, addresses vendor independence, a key concern for BCS practitioners). Structurally, TRA strongly predicts EOU ($\beta = 0.79$, $p < .001$), with EOU the dominant driver of USF ($\beta = 1.00$, $p < .01$), and USF significantly predicts USE ($\beta = 0.74$, $p < .05$). The model explains 62\%, 69\%, and 44\% of the variance in EOU, USF, and USE, respectively. The non-significant EOU $\rightarrow$ USE path is consistent with classic TAM findings~\cite{davis_user_1989}, where EOU affects intention primarily through USF rather than directly.

Overall, the results support moderate-to-high acceptance of our proposed solution, consistent across civil engineers and computer scientists alike, including those with no prior DSML or PLC experience. We discuss implications in \cref{sec:discussion}.

\subsection{Technical Performance}

To validate R5 and answer \textbf{RQ2}, we evaluated generation time and end-to-end deployment time across models of varying complexity. All tests ran on a laptop with Windows 11 Home, an AMD Ryzen 5 6600HS processor, and 16\,GB of memory.

\paragraph{Generation Performance}

We varied three dimensions: hardware datapoints (10--1000), control logic statements  (50--5000), and nesting depth (up to 5 levels). \cref{tab:generation} reports our results. Generation time scales primarily with statement count and datapoint count, whereas nesting depth has a comparatively smaller effect at low statement counts but compounds significantly at 5000 statements. The largest model tested (1000 datapoints, 5000 statements, 5 nesting levels) completed in \qty{19.9}{\second}, confirming suitability for iterative development on large-scale projects.
\begin{table*}[bth]
    \begin{minipage}[t]{0.48\linewidth}
        \centering
        \scriptsize
        \caption{Generator performance by model sizes. Times are in \unit{\second}.}
        \label{tab:generation}
        \begin{tabular}{
          l
          S[table-format=2.1]
          S[table-format=2.1]
          S[table-format=2.2]
          S[table-format=2.1]
          S[table-format=2.1]
          S[table-format=2.2]
          S[table-format=2.1]
          S[table-format=2.1]
          S[table-format=2.1]
        }
            \toprule
            Datapoints & \multicolumn{3}{c}{10} & \multicolumn{3}{c}{250} & \multicolumn{3}{c}{1000} \\
            Nesting    & {none} & {3} & {5} & {none} & {3} & {5} & {none} & {3} & {5} \\
            Statements & \multicolumn{9}{c}{} \\
            \midrule
            50   & 0.9  & 1.0  & 1.1  & 1.1  & 1.2  & 1.3  & 1.1  & 1.2  & 1.3  \\
            1000 & 1.2  & 1.6  & 1.8  & 1.9  & 4.5  & 2.7  & 3.8  & 4.1  & 4.6  \\
            5000 & 2.1  & 3.7  & 4.7  & 4.8  & 6.6  & 7.7  & 15.2 & 18.7 & 19.9 \\
            \bottomrule
        \end{tabular}
    \end{minipage}\hfill
    \begin{minipage}[t]{0.49\linewidth}
        \centering
        \scriptsize
        \caption{End-to-end deployment time by scenario. Times are in \unit{\second}.}
        \label{tab:deployment}
        \begin{tabular}{lS[table-format=2.1]S[table-format=2.1]S[table-format=2.1]}
            \toprule
            {Scenario} & {Mean} & {Min} & {Max} \\
            \midrule
            Baseline                        & 35.7 & 34.1 & 37.3 \\
            High Logic Complexity           & 34.9 & 33.2 & 36.8 \\
            High Declaration Complexity     & 38.9 & 37.4 & 40.6 \\
            High Type System Complexity     & 41.7 & 40.1 & 43.5 \\
            Max Complexity                  & 47.1 & 45.3 & 49.2 \\
            \bottomrule
        \end{tabular}
    \end{minipage}
\end{table*}

\paragraph{Deployment Performance}

We held the hardware model constant and varied control model complexity across five scenarios on the physical testbench. Each scenario targets a different dimension of model complexity. \emph{Baseline} uses minimal types, function blocks, and logic. \emph{High Logic Complexity} adds 250 \texttt{use} statements. \emph{High Declaration Complexity} adds 50 function block declarations. \emph{High Type System Complexity} adds 50 structs and 50 enums. \emph{Max Complexity} combines all factors simultaneously. As shown in \cref{tab:deployment}, the total deployment time is dominated by the fixed overhead of the Beckhoff Automation Interface. The baseline deployed in \qty{35.7}{\second}, and the worst case reached \qty{47.1}{\second}. All scenarios completed in under \qty{1}{\minute}, confirming that the framework introduces no prohibitive bottlenecks even for complex configurations.

\section{Discussion}
\label{sec:discussion}

Our results support the viability of our proposal. All participants completed the three tasks within 45 minutes, compared to roughly 3 hours for the manual baseline, with the DSML consistently producing 31--37\% fewer lines of code across all scenarios. Adaptation tasks were completed in under 15 minutes compared to approximately 30 minutes manually, which yielded a $2\times$ speedup. The scalability results show acceptable generation times even for large models, and deployment is consistently completed under 1 minute. User acceptance was moderate-to-high across all participants, including those with no prior DSML or PLC experience. These results align with reported MDE benefits~\cite{whittle2013state,BUSCHHAUS2024103033}, and extend them into the AECO domain. W.r.t.~related
work~\cite{knorr2024model,schachinger2015model,cai2023automated}, our approach is the first to provide a fully automated, end-to-end process pipeline from platform-independent models to deployed BCS without manual
bottlenecks in the transformation or deployment steps. Where prior approaches either stop at simulation artifacts~\cite{stratbucker2017model} or require manual tuning after generation~\cite{cai2023automated}, our
pipeline runs without intervention from model to live system.

We currently target Beckhoff TwinCAT, but our design is platform-independent. Both DSMLs operate at a level of abstraction decoupled from any
vendor-specific runtime, and the code generator is the only platform-specific component. Swapping it for a generator targeting a different PLC runtime, such as CODESYS (e.g., for WAGO PLC) or TIA Portal (e.g., for Siemens PLC), requires no changes to the models or methodology. Since both DSMLs emit standard IEC~61131-3 ST~\cite{iec61131-3}, extending the pipeline to other ST-compatible
platforms is straightforward. This positions our framework as a vendor-independent foundation for model-driven BCS engineering by addressing a crucial gap identified in the literature~\cite{hammes2024market,gotz2020model}.

For \textbf{RQ1}, our results demonstrate that the combination of a dedicated \texttt{Hardware DSML} and \texttt{Control DSML}, together with model-driven generation, provides a viable abstraction layer that decouples control logic from vendor-specific implementation while supporting automated end-to-end deployment without manual intervention. For \textbf{RQ2}, we demonstrated significant reductions in both engineering time and lines of code, and a roughly 2$\times$ speedup for adaptation tasks compared to the manual baseline, with gains holding across participants with varying levels of prior experience. For \textbf{RQ3}, participants with no prior DSML or PLC knowledge completed all modeling tasks successfully, and the TAM study confirms moderate-to-high acceptance across all participant groups. The strong EOU $\rightarrow$ USE path indicates that once participants found the DSML easy to learn, they also perceived it as useful, which is a meaningful signal for practical adoption in the AECO domain. Taken together, these results suggest that a model-driven process is both viable and accessible for BCS engineering in practice.

\subsection{Implications \&  Lessons Learned}

For researchers, this work shows that the AECO domain is a viable and largely underexplored application area for MDE~\cite{gotz2020model}. The strong EOU $\rightarrow$ USF path and high values for USF suggest that domain practitioners can effectively adopt our DSMLs without a deep SE background, which broadens the scope of MDE applicability for process improvement.

For practitioners, the key takeaway is that the barrier to BCS programming can be substantially lowered. Participants with no prior DSML or PLC experience were able to complete all three tasks (cf.~\cref{ssec:efficiency}). This has direct implications for facility managers and specialist planners, who today must re-engage costly external contractors for even minor BCS updates. The DSML-based process pipeline reduces that dependency by making the system directly modifiable at a high level of abstraction.

Several lessons emerged from building and evaluating the framework. The separation of hardware and control modeling was the single most sustainable design decision, as it allowed participants to reason about physical and logic changes independently, which directly contributed to the fast adaptation times in EQ2.

The design of the \texttt{Control DSML} was perceived to be largely intuitive and easy to use. Participants were able to complete all tasks without major complications. During our study, we observed that core elements, such as units and function blocks, were consistently well understood and could be applied without extensive prior experience. However, several participants found the \texttt{use} syntax for invoking function blocks unintuitive and harder to read at first. One participant suggested that \emph{``predefined templates or some available examples could have helped to make this easier''}, which could also be complemented by more advanced training. Furthermore, tooling integration also matters more than expected. Syntax highlighting, code completion, and on-the-fly validation via Langium~\cite{eclipse_langium} within VS Code significantly reduced cognitive load, particularly for participants unfamiliar with the domain. Several participants noted that inline feedback helped them recover from errors without consulting documentation.

Finally, the Beckhoff Automation Interface introduced a fixed deployment overhead of roughly 35~s. This is acceptable for commissioning workflows but may be problematic for rapid iterative development. An iterative approach, i.e., redeploying only what was changed, would improve commissioning times and user experience.

\subsection{Limitations and Threats to Validity}
\label{ssec:threats}

As with any empirical study, our results are subject to several threats to validity. As to \emph{internal validity}, the manual baseline was implemented by one of the authors, which may introduce bias due to domain familiarity. Participants also had limited time to explore the tooling and problem domain. To mitigate this, we provided a detailed DSML description, a tool introduction, and task guidelines prior to the study.

For \emph{construct validity}, our efficiency metrics, viz.~time-on-task, lines of code, and number of artifacts modified, may not fully capture all aspects of engineering effort. We applied GQM~\cite{caldiera1994goal} to select metrics that directly address our research questions.

As to \emph{external validity}, we conducted our study with $n=15$ participants in a controlled setting using a single hardware platform. While the scenarios reflect realistic BCS engineering tasks, a larger-scale industrial case study with more participants and broader hardware coverage is needed to fully validate generalizability.

Finally, given $n=15$, the estimated EOU~$\rightarrow$~USF coefficient of $\beta = 1.00$ lies at the boundary of the parameter space and should be read as exploratory rather than as a literal effect size. Our sample further contains no practicing \emph{facility manager}, although \cref{fig:solution} assigns that role the post-commissioning adaptation loop, so S3 is only a proxy for it. Finally, our DSMLs cover the constructs required by the scenarios studied; a genuinely missing abstraction would require changing grammar, validator, and generator templates together, which we regard as a real maintenance cost of the approach. Since the same participants perform tasks and immediately afterwards evaluate the solution they produced, judgments regarding performance and acceptance may correlate without necessarily reflecting long-term usage intentions.

\section{Conclusion}
\label{sec:conc}

We presented a model-driven process pipeline combining two platform-independent DSMLs with automated code generation and fully automated deployment to an operational BCS installation. 
Our evaluation confirms the approach works in practice. The DSML produced 31--37\% fewer lines of code and code generation scales to large models. End-to-end deployment is consistently completed in under 1 minute. 
These results suggest that a model-driven workflow with a defined, role-assigned engineering process is viable for improving BCS engineering in practice. The platform-independent design of our DSMLs makes it straightforward to retarget our pipeline to other IEC~61131-3 ST-compatible and open platforms (though necessary remote APIs need to be available for automated deployment), such as CODESYS (e.g., for WAGO PLC) or TIA Portal (e.g., for Siemens PLC), by adapting the code generator. Our solution artifact is available for download~\cite{artifact} for experimentation. Future work includes a larger-scale industrial study with more participants and broader hardware coverage, integration of a hot-reload mechanism to reduce deployment overhead for rapid iterative changes, and exploration of models@run.time support for continuous adaptation during building operation~\cite{nelly19}.


\section*{Acknowledgments}
This research has received funding from the Austrian Research Promotion Agency (FFG) under Grant Agreement No.:~898708, TwinLight.




\balance

\bibliographystyle{abbrv}
\bibliography{references}

\end{document}